\documentclass[12pt]{iopart}
\usepackage{graphicx}
\usepackage{amssymb}
\usepackage{dcolumn}
\usepackage{citesort}
\usepackage{iopams}
\usepackage{datetime}
\usepackage[margin=0cm]{caption}
\begin{document}

\title{Penetration depth study of the type-I superconductor PdTe$_2$}

\author{M. V. Salis$^1$, P. Rodi\`{e}re$^2$, H. Leng$^1$, Y. K. Huang$^1$, A. de Visser$^1$}

\address{$^1$ Van der Waals - Zeeman Institute, University of Amsterdam, Science Park 904, 1098 XH Amsterdam, The Netherlands}
\address{$^2$ Institut N\'{e}el, CNRS \& Universit\'{e} Grenoble Alpes, BP 166, 38042 Grenoble Cedex 9, France}
\eads{\mailto{m.v.salis@uva.nl}, \mailto{a.devisser@uva.nl}}

\begin{abstract}
Superconductivity in the topological non-trivial Dirac semimetal PdTe$_2$ was recently shown to be type-I. We here report measurements of the relative magnetic penetration depth, $ \Delta \lambda$, on several single crystals using a high precision tunnel diode oscillator technique. The temperature variation $\Delta \lambda (T)$ follows an exponential function for $T/T_c < 0.4$, consistent with a fully-gapped superconducting state and weak or moderately coupling superconductivity. By fitting the data we extract a $\lambda (0)$-value of $\sim 500$~nm. The normalized superfluid density is in good agreement with the computed curve for a type-I superconductor with nonlocal electrodynamics. Small steps are observed in $\Delta \lambda (T)$, which possibly relates to a locally lower $T_c$ due to defects in the single crystalline sample.

\vspace{0.8cm}
\end{abstract}
\maketitle
\section{Introduction}

Transition metal dichalcogenides display a wide variety of interesting electronic properties. Recently, these layered materials received considerable attention because they offer a fruitful playground for the realization of topological non-trivial electronic band structures~\cite{Soluyanov2015,Huang2016,Yan2017,Bahramy2018}. Notably, it has been proposed that transition metal dichalcogenides exhibit a generic coexistence of type-I and type-II three-dimensional Dirac fermion states~\cite{Bahramy2018}. A prominent example is PdTe$_2$ that was reported to be a type-II Dirac semimetal, based on \textit{ab-initio} electronic structure calculations combined with angle-resolved photoemission spectroscopy (ARPES)~\cite{Liu2015a,Fei2017,Noh2017,Bahramy2018,Clark2017}. In a type-II Dirac semimetal the Dirac cone is tilted and Lorentz invariance is broken~\cite{Soluyanov2015}. The Dirac point then forms the touching point of electron and hole pockets. PdTe$_2$ is of special interest because it is a superconductor as well ($T_c = 1.6$~K~\cite{Guggenheim1961}). Especially, it has been proposed to be an improved platform for topological superconductivity~\cite{Fei2017}. Topological non-trivial superconductors attract a strong interest since they are predicted to host protected Majorana zero modes at their surface (for recent reviews see references \cite{Sato&Fujimori2016,Sato&Ando2017}). This offers a unique design route for devices based on topological quantum computation.\\

Recently, the superconducting phase of PdTe$_2$ was characterized in detail by magnetic and transport experiments by Leng \textit{et al.}~\cite{Leng2017}. Surprisingly, dc-magnetization measurements on high-quality single crystals show that PdTe$_2$ presents a rare case of a binary compound that is a type-I superconductor. The crystals also show the intermediate state as demonstrated by the differential paramagnetic effect that is observed in low frequency ac-susceptibility measurements in applied dc-fields. The magnetization data reveal superconductivity is a bulk property with $T_c = 1.64$~K and a critical field $H_c (T)$ that follows a quadratic temperature variation with $\mu_0 H_c (0) = 13.6$~mT. This provides strong evidence that PdTe$_2$ is a conventional type-I Bardeen-Cooper-Schrieffer (BCS) superconductor. Interestingly, in applied fields $H_a > H_c$ large diamagnetic screening signals persist in the ac-susceptibility measured for small driving fields~\cite{Leng2017}. This is attributed to superconductivity of the surface sheath. However, the critical field for surface superconductivity $H_c ^S$ does not obey the standard Saint-James - de Gennes relation with critical field $H_{c3} = 2.39 \times \kappa H_c$~\cite{Saint-James&deGennes1963}, where $\kappa$ is the Ginzburg-Landau parameter. Therefore, it was proposed~\cite{Leng2017} superconductivity of the surface layer has a topological nature and originates from the topological surface states that were detected by ARPES~\cite{Liu2015a,Noh2017}.\\

The superconducting properties of PdTe$_2$ were further investigated by heat capacity measurements~\cite{Amit2018}, as well as by scanning tunneling microscopy (STM) and spectroscopy (STS)~\cite{Das2018,Clark2017}. The specific heat data confirm conventional BCS superconductivity with a ratio $\Delta C/ \gamma T_c \approx 1.52$, which is close to the weak-coupling value 1.43. Here $\Delta C$ is the size of the step in the specific heat at $T_c$ and $\gamma$ the Sommerfeld coefficient. The zero-field STM/STS measurements point to a fully-gapped superconducting state. The absence of in-gap states seems to rule out topological superconductivity at the surface~\cite{Das2018,Clark2017}. However, the spectra in applied fields differ in important details. The conductance images taken by Das \textit{et al.}~\cite{Das2018} are consistent with type-I superconductivity with a critical field of 25~mT, but at certain points on the surface much larger critical fields (up to 4~T) were observed. This was attributed to the presence of randomly distributed impurities/defects. We remark these higher fields are reminescent of the transport critical field $\mu_0 H_c ^R (0) =0.3$~T $ > \mu_0 H_c ^S (0)$ reported by Leng \textit{et al}.~\cite{Leng2017}. On the other hand, Clark \textit{et al}.~\cite{Clark2017} claim to observe a vortex, indicating the superconductivity they probe is type-II. We remark, the presence of the type-I intermediate state that arises due to the sizeable non-zero demagnetization factor, $N$, has not been considered in the interpretation of the tunneling experiments. We conclude these conflicting results demand for a further examination of superconductivity in the bulk and on the surface of PdTe$_2$.\\

We here present measurements of the relative magnetic penetration depth, $\Delta \lambda$, of PdTe$_2$ single crystals using the tunnel diode oscillator technique. The $\Delta \lambda (T)$-variation shows an exponential temperature dependence for $T/T_c < 0.4$ consistent with a fully-gapped superconducting state. The $\lambda (0)$-value extracted by fitting $\Delta \lambda (T)$, measured for several crystals, to the standard BCS expression amounts to 500~nm. The normalized superfluid density is in good agreement with the computed curve for a type-I superconductor with nonlocal electrodynamics and $\lambda (0) = 121$~nm.

\section{Experiment}

PdTe$_2$ crystallizes in the trigonal CdI$_2$ structure (space group P$\bar{3}$m1)~\cite{Thomassen1929}. The crystals used in this work were cut from the crystal grown by a modified Bridgman technique~\cite{Lyons1976} as reported and characterized in reference \cite{Leng2017}. Scanning Electron Microscopy (SEM) with Energy Dispersive X-ray (EDX) spectroscopy showed the proper 1:2 stoichiometry within the experimental resolution of 0.5~\%. Powder X-ray diffraction confirmed the CdI$_2$ structure. From transport measurements a residual resistance ratio $R$(300K)/$R$(2K) = 30 was deduced. Magnetic measurements show $T_c = 1.64$~K and $\mu _0 H_c(0) = 13.6$~mT~\cite{Leng2017} (for an applied field directed in the plane of the layers, $H_a \parallel a$-axis). Three crystals (labeled s1, s2 and s3) with approximate size $1.0 \times 1.0 \times 0.1$~mm$^3$ were cut from the bulk crystal using a scalpel blade. The thin direction is taken along the $c$-axis. After the measurements the thickness of crystal s3 was reduced to $\sim 7~\mu$m by the Scotch tape technique. This crystal is labeled s4. The dimensions of the crystals are listed in table \ref{tab:table1}.\\

Magnetic penetration depth measurements were carried out with the tunnel diode oscillator (TDO) technique \cite{Degrift1975}. Changes in the resonant frequency $f$ of a high stability $LC$ circuit are measured with a tunnel diode as gain operating at $\sim$ 13.4~MHz~\cite{Diener2008}. The ac-excitation field, $H_{ac}$, is below 1~$\mu$T and the earth magnetic field is shielded by mu-metal. The sample is fixed with vacuum grease at the end of a sapphire rod attached to the cold finger of a home-built helium-3 refrigerator, which allows for measurements down to 0.38~K~\cite{Diener2008}. During measurements the sample is positioned in the center of the excitation coil with inductance $L$. The crystal can be extracted \textit{in-situ} from the excitation coil for calibration purposes. Changes in the penetration depth with respect to temperature are reflected in the resonant frequency via the mutual inductance of the sample and the coil. The frequency shift $\Delta f$ is measured relative to the frequency at the base temperature of the refrigerator. Measurements were made on crystals s1-s4 mounted vertically, \textit{i.e.} with the field applied in the plane of the layers, $H_{ac} \perp c$, such that the screening currents flow in the plane and along the $c$-axis. In this case the frequency shift $\Delta f$ can be converted to a relative shift of $\lambda$ according to~\cite{Fletcher2007}:

\begin{equation}\label{eq:freqtolambda}
\frac{\Delta f}{\Delta f_0} = \frac{2}{t} \bigg(\Delta\lambda_{a} + \frac{t}{w} \Delta\lambda_c \bigg).
\end{equation}

Here $\Delta \lambda _a$ and $\Delta \lambda _c$ are the relative magnetic penetration depths along the $a$- and $c$-axis, respectively, and $t$ and $w$ are the thickness and width of the crystal, respectively. $\Delta f_0$ is the frequency shift obtained by extracting the sample from the coil and accounts for demagnetization effects and the coil calibration factor. For thin square sample geometries the latter term in the brackets can be neglected since $w \gg t$. Crystal s1 was also measured in the horizontal configuration, with $H_{ac} \parallel c$, and the screening currents in the $aa^*$-plane. For a thin square geometry, with the magnetic field applied perpendicular to the broad surface, $\Delta f$ can be converted to the in-plane relative magnetic penetration depth $\Delta \lambda _a$ according to \cite{Prozorov2000}:
\begin{equation}
\Delta \lambda _a = -R \frac{\Delta f}{\Delta f_0},
\label{eq:hcfreqtolambda}
\end{equation}
where $R$ is a characteristic geometrical factor given by
\begin{equation}
R = \frac{w/2}{
2 \bigg[ 1 + \bigg[ 1 + \bigg( \frac{2t}{w} \bigg)^2 \bigg] \textrm{arctan} \bigg( \frac{w}{2t} \bigg) - \frac{2t}{w} \bigg] },
\end{equation}
with $t$ the sample thickness and $w$ the mean of the in-plane dimensions.


\begin{table}
\caption{Dimensions (length, width and thickness) and onset superconducting transition temperature, $T_c^{onset}$, of the PdTe$_2$ crystals investigated.\\ }
\begin{center}
\begin{tabular}{c c c c c}
\hline
\hline
Crystal & ~~length~~&~~width~~&~~thickness~~&~~$T_c^{onset}$~~\\
& (mm) & (mm) & (mm) & (K)\\
\hline
s1&0.88&0.84&0.097&1.67\\
s2&1.20&1.15&0.096&1.65\\
s3&1.48&0.96&0.060& 1.63\\
s4 &1.18&0.88&$\sim 0.007$&1.76\\
\hline
\hline
\end{tabular}
\end{center}
\label{tab:table1}

\end{table}

\section{Results and analysis}

\begin{figure}[ht]
\centering
\includegraphics[width=10cm]{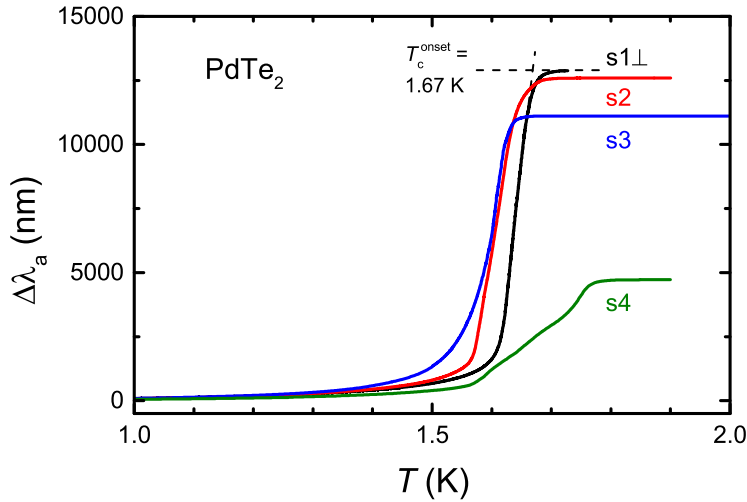}
\caption{Temperature dependence of $\Delta \lambda _a$ of four PdTe$_2$ crystals labeled s1-s4 with $H_{ac} \perp c$. $\Delta \lambda _a$ is computed from the frequency shift $\Delta f$ according to equation \ref{eq:freqtolambda}. Dashed lines for crystal s1 show how $T_c ^{onset} = 1.67$~K is determined. In labeling crystal s1 we added the symbol $\perp$ since this crystal was measured in two configurations (see text). }
\label{fig:figure1}
\end{figure}

\begin{figure}[ht]
\centering
\includegraphics[width=10cm]{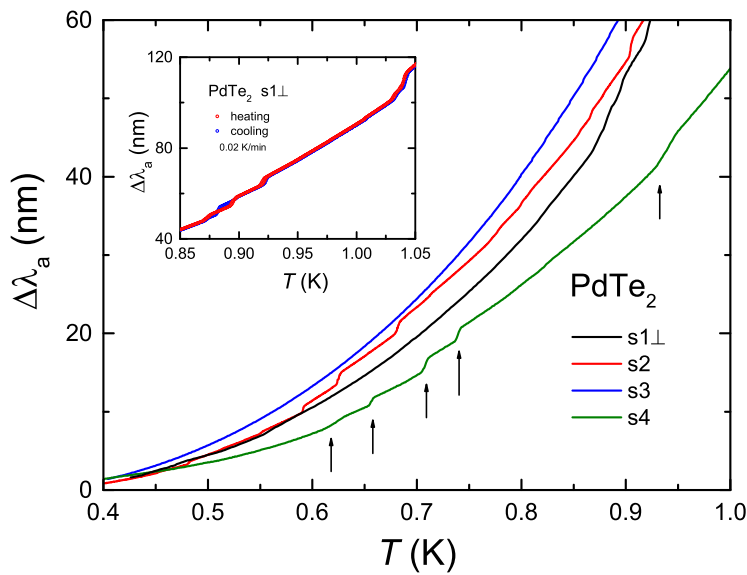}
\caption{Zoom-in of the data presented in figure \ref{fig:figure1} for 0.4~K~$ < T < $~1~K. Small steps are visible in $\Delta \lambda _a$ (indicated by arrows), notably for crystals s2 and s4. The inset shows data taken on heating (red symbols) and cooling (blue symbols) at a rate of 0.02~K/min, which demonstrates that the steps are reproducible with little hysteresis.}
\label{fig:figure2}
\end{figure}

\begin{figure}[ht]
\centering
\includegraphics[width=10cm]{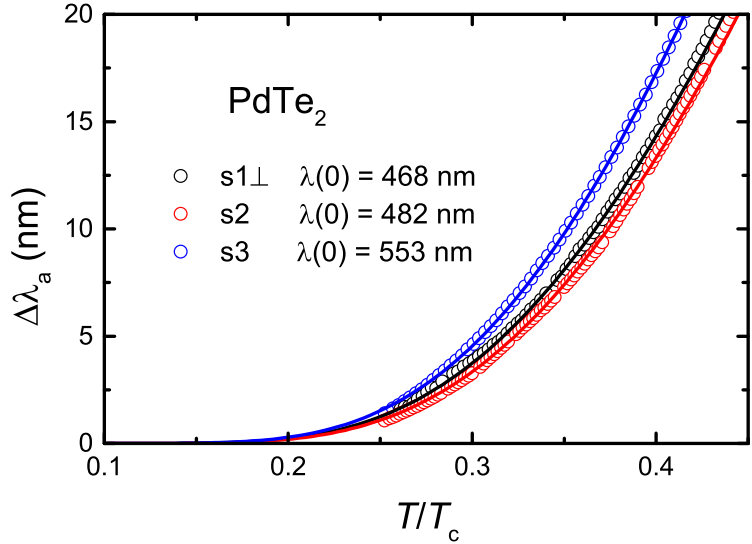}
\caption{$\Delta \lambda_a$ as a function of the reduced temperature $T/T_c$ for PdTe$_2$ crystals s1-s3 measured for $H_{ac} \perp c$. Open circles: experimental data, with small steps removed for crystals s1 and s2 (see text). Solid lines: BCS fits using equation \ref{eq:bcsfit} up to $T/T_c = 0.4$. }
\label{fig:figure3}
\end{figure}

The overall temperature variation of the in-plane relative magnetic penetration depth of all four crystals is reported in figure \ref{fig:figure1}. In this experiment the crystals were mounted vertically and $\Delta \lambda_a (T)$ was extracted using equation \ref{eq:freqtolambda}. For crystals s1-s3 the superconducting transition is sharp, with $T_c ^{onset}$-values in the range 1.63-1.67~K, while for the thin crystal (s4) the transition is broad with $T_c ^{onset} = 1.76$~K. Above $T_c$ the tunnel diode oscillator technique probes the skin depth $\delta$, which is temperature independent in this case since the electrical resistivity, $\rho(T)$, attains a constant value near $T_c$~\cite{Leng2017,Hardy1993}. In the measurements on crystals s1, s2 and s4 tiny steps in $\Delta \lambda_a$ are observed as shown in figure \ref{fig:figure2}. The steps have different sizes up to 3~nm and appear both upon cooling and warming and are reproducible (see inset \ref{fig:figure2}). Smoothed data sets for crystal s1 and s2 are readily obtained after removing the steps as shown in figure \ref{fig:figure3}. In the following paragraphs we analyze $\Delta \lambda_a (T)$ as measured and with the steps removed. We remark the results differ in details only, and the presence or absence of steps does not affect any of our main conclusions.\\

The gap symmetry of the superconducting state can be elucidated by examining the temperature dependence of $\lambda _a$, or $\Delta \lambda _a$. For a superconductor with an isotropic superconducting gap, $\Delta \lambda (T)$ at low temperatures follows the relation~\cite{Poole2007}

\begin{equation} \label{eq:bcsfit}
\frac{\Delta \lambda(T)}{\lambda(0)} = \sqrt{\frac {\pi \Delta(0)}{2 k_B T}} e^{- \frac{\Delta(0)}{k_B T}}.
\end{equation}

Here $\lambda(0)$ and $\Delta(0)$ are the penetration depth and superconducting gap for $T \rightarrow 0$, respectively, and $k_B$ is the Boltzmann constant.
For a weak-coupling BCS superconductor $\Delta(0) = 1.76k_B T_c$. We have fitted $\Delta \lambda_a$ to equation \ref{eq:bcsfit} with $\lambda_a(0)$ and $\Delta(0)$ as fit parameters, where a small offset has been added as a fit parameter to $\Delta \lambda_a$ to account for the exponential variation between $\lambda_a(0)$ and $\lambda_a(0.38$K$)$. It appears good fits can be obtained when we restrict the temperature range to $T/T_c < 0.4$ as shown in figure \ref{fig:figure3} for crystals s1-s3. In table \ref{tab:table2} we summarize all fit parameters obtained by fitting the as-measured and smoothed data for the four crystals. The value for $\lambda_a(0)$ ranges between 468 and 586~nm, with an outlier of 745~nm for the as-measured crystal s2. The latter large value is attributed to the ubiquitous presence of steps in the fitted range. The ratio $\Delta(0) /k_B T_c$ ranges from 1.77 to 1.86, again with the value 1.97 as an outlier for the as-measured crystal s2. These values are close to the weak-coupling BCS value 1.76. The error bar on the fit parameters is of the order of 10~\%, a value obtained by inspecting the stability of the fit when slightly changing the reduced temperature range, $0.35 < T/T_c < 0.45$. The values of the fit parameters listed in table \ref{tab:table2} show smoothing the data does not bring about major changes, the main reason being that the steps in the fitted temperature range are very small (except for crystal s2).

\begin{table}
\caption{Fit parameters $\Delta(0)/k_BT_c$ and $\lambda(0)$ obtained by fitting equation \ref{eq:bcsfit} to the as-measured PdTe$_2$ data (upper block) and data with steps removed (lower block). Entries for crystals s2-s4 are for $H_{ac} \perp c$-axis and for crystal s1 with $H_{ac} \perp c$-axis and $\parallel c$-axis.\\}
\begin{center}
\begin{tabular}{c c c c c}
\hline
\hline
Crystal & ~~range~~ & ~~$\Delta(0)/k_BT_c$ ~~&$\lambda(0)$ \\
measured data & ($T/T_c$)& &(nm)\\
\hline
s1$\perp$ & 0.4&1.77&471 \\
s2&0.5&1.97&745\\
s3&0.4&1.78&553\\
s4&0.5&1.86&586\\
\hline
smoothed data\\
\hline
s1$\perp$&0.4&1.77&468\\
s1$\parallel$&0.4&1.74&377\\
s2&0.4&1.83&482\\
s3&-&-&-\\
s4&0.5&1.83&471\\
\hline
\hline
\end{tabular}
\end{center}
\label{tab:table2}
\end{table}

\begin{figure}[ht]
\centering
\includegraphics[width=10cm]{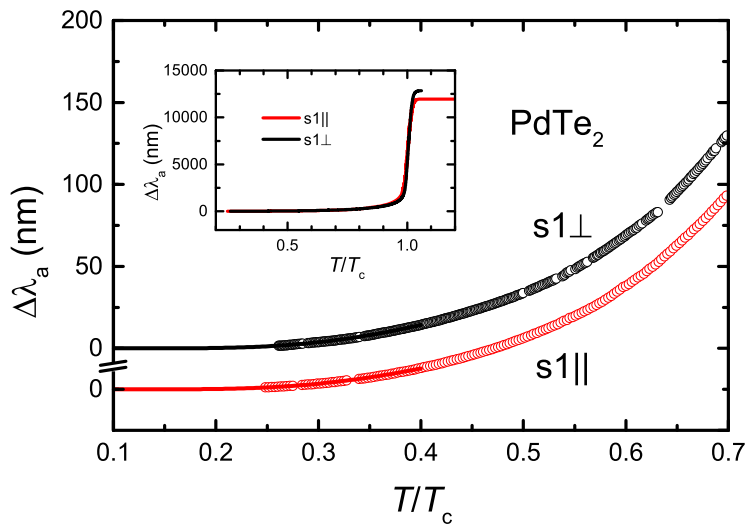}
\caption{Temperature dependence of $\Delta\lambda_a$ for crystal s1 with the driving field perpendicular (black circles) and parallel to the $c$-axis (red circles), where small steps have been removed for both data sets. The solid lines represent fits to equation \ref{eq:bcsfit} up to $T/T_c = 0.4$. Inset: Temperature dependence of $\Delta \lambda_a$ in a larger temperature range.}
\label{fig:figure4}
\end{figure}

In figure \ref{fig:figure4} we report $\Delta \lambda_a (T)$ of crystal s1 with the excitation field applied parallel to the $c$-axis. Here $\Delta \lambda_a (T)$ is calculated by using equation \ref{eq:hcfreqtolambda}. In this configuration a few tiny steps appeared in the as-measured data as well, but less frequently than for the same crystal in the perpendicular configuration. In figure \ref{fig:figure4} these steps have been removed. A fit to equation \ref{eq:bcsfit} up to $T/T_c = 0.4$ yields $\lambda_a(0) = 377 $~nm and $\Delta(0)/k_B T_c = 1.74$. These fit parameters are listed in table \ref{tab:table2} as well. For comparison we show in figure \ref{fig:figure4} $\Delta \lambda_a (T)$ measured for $H_{ac} \perp c$ on the same crystal. Overall, the data are in good agreement, as well as the extracted values of $\lambda_a(0)$ given a 10\% error margin. We remark that the calculated value of $\lambda_a(0)$ for $H_{ac} \perp c$ would be slightly reduced if we do not neglect the second term in equation \ref{eq:freqtolambda}.

\section{Superfluid density}

To further elucidate the nature of the superconducting phase of PdTe$_2$, especially the type-I or type-II character, we analyze $\Delta \lambda (T)$ (from now on we drop the subscript a) in terms of the normalized superfluid density $\rho_s(T)$~\cite{Poole2007}

\begin{equation} \label{eq:nrho}
\rho_s (T) = \frac{n_s(T)}{n_s(0)} = \frac{\lambda (0)^2}{\lambda (T)^2} =\Bigg( \frac{\lambda(0)}{\lambda(0) + \Delta \lambda(T)}\Bigg)^2.
\end{equation}

\begin{figure}[ht]
\centering
\includegraphics[width=12cm]{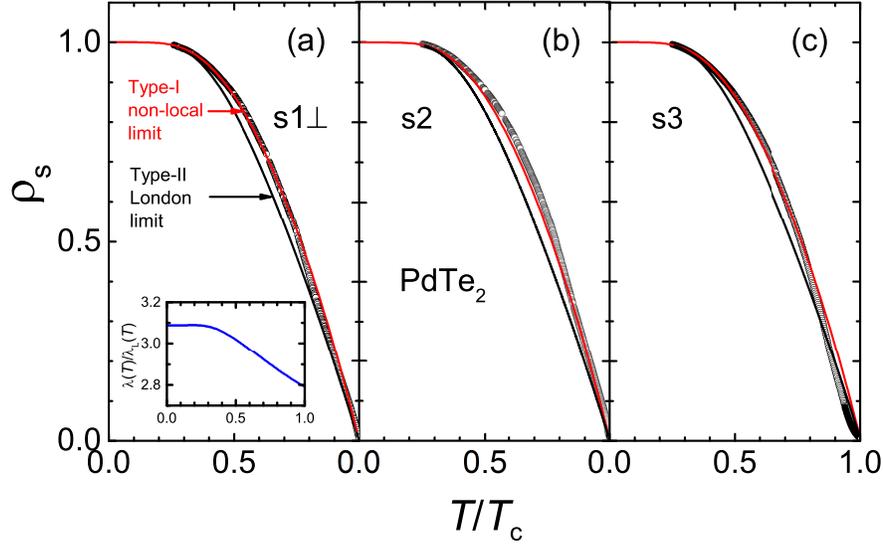}
\caption{(a) Normalized superfluid density $\rho_s(T)$ for PdTe$_2$. Open circles: calculated values for crystal s1 using equation \ref{eq:nrho} with $\lambda (0) = 468$~nm. Red solid line: computed $\rho_s$ for the nonlocal case with help of equations \ref{eq:nrho} and \ref{eq:miller} with parameters $\lambda_L(0) = 39$~nm, $\xi(0)=$~1.8~$\mu$m and $\ell= 531$~nm (see text). Black solid line: computed $\rho_s$ in the London limit with help of ~8 and parameter $\Delta(0)/k_BT_c = 1.77$. Inset: Computed temperature variation of the ratio $\lambda(T) / \lambda_L(T)$. (b) and (c) The same analysis for crystals s2 and s3 with $\lambda (0) = 482$~nm and 553~nm, respectively.  }
\label{fig:figure5}
\end{figure}

We first trace in figure \ref{fig:figure5}a the normalized superfluid density of crystal s1 ($H_{ac} \perp c$) calculated from the temperature variation of $\Delta \lambda (T) $, where we used $\lambda (0) = 468$~nm (see table \ref{tab:table2}). Here the reduced temperature $T/T_c$ is calculated with $T_c = 1.63$~K, which locates the midpoint of the superconducting transition, rather than with $T_c^{onset} = 1.67$~K. Next we compare the experimental curve to theoretical expressions for $\rho _s$ evaluated with (type-I) and without (type-II) nonlocal electrodynamics.

For a type-I superconductor nonlocal electrodynamics has to be taken into account, because the field penetration depends on the ratio of the BCS coherence length $\xi_0$ and the electron mean free path $\ell$. This results in an enhancement of the ratio $\lambda(T) / \lambda_L(T)$, where $\lambda_L(T)$ is the London penetration depth and $\lambda(T)$ is the magnetic penetration depth due to nonlocal electrodynamics~\cite{Miller1959}. In the limit $T \rightarrow 0$, the London penetration depth is given by $\lambda_L (0) = (m^*/\mu_0 n_s e^2)^{1/2}$, where $m^*$ is the effective mass, $n_s$ the superfluid density, $\mu_0$ the vacuum permeability and $e$ the elementary charge. With a carrier density $n=5.5 \times 10^{27}$~m$^{-3}$ \cite{Leng2017} and $ m^* \approx 0.3m_e$~\cite{Dunsworth1975} (here we use an average value $ m^*$ and $m_e$ is the free electron mass) a value of $\lambda _L(0)$ of 39~nm had been derived~\cite{Leng2017}. Considering Pippard's nonlocal electrodynamics~\cite{Tinkham1996} the penetration depth is enhanced by the ratio $\lambda(0) / \lambda_L(0)= (1+\xi_0 / \ell)^{1/2}$ in the limit $T \rightarrow 0$. The BCS coherence length is given by $\xi_0 = \hbar v_F / \pi \Delta(0) = 0.18 \hbar^2 k_F /(k_B T _c m^*)$ with $v_F$ the Fermi velocity and $k_F$ the Fermi wave number. With $k_F = (3 \pi ^2 n) ^{1/3} = 5.5 \times 10^9$~m$^{-1}$ we calculate $\xi_0 = 1.8$~$\mu$m. We remark $\xi_0$ is larger than the coherence length $\xi = 439$~nm derived using Ginzburg-Landau relations\footnote{In reference \cite{Leng2017} the calculation of the coherence length $\xi$ contains an error. The correct value is 439~nm}. An estimate for $\ell$ can be taken from the Drude transport model $\ell = 3 \pi^2 \hbar/\rho_0 e^2 k_F ^2$. With the residual resistivity $\rho_0 = 0.76~ \mu \Omega$cm~\cite{Leng2017} we calculate $\ell = 531$~nm. Consequently $\xi_0 / \ell \approx 3.4$ and $\lambda(0) / \lambda_L(0) = 2.1$.

A full fledged function $\lambda(T) / \lambda_L(T)$ in the case of nonlocal electrodynamics can be computed following Miller~\cite{Miller1959}, where

\begin{equation}\label{eq:miller}
\lambda = \pi \Bigg[ \int_0^{\infty} \textrm{ln}\bigg[ 1 + q^{-2} K(q) \bigg]dq \Bigg]^{-1}.
\end{equation}

Here the response function $K(q)$ can be calculated with help of equations 5.26 and 5.3 in reference \cite{Nam1967}.
Using the parameters $\lambda_L(0) = 39$~nm, $\xi_0 = 1.8$~$\mu$m and $\ell = 531$~nm derived above, we have computed $\rho_s (T)$ and the result is shown as the solid red line in figure \ref{fig:figure5}a. In the inset we show the calculated ratio $\lambda(T) / \lambda_L(T)$. We remark that for temperatures below $T/T_c = 0.4$ the ratio is constant within 1~\%, indicating that at low temperatures the local and nonlocal cases are indistinguishable, apart from the absolute value of the ratio. At higher temperatures the ratio is no longer constant, which allows one to discriminate between the local and nonlocal case via the superfluid density. In the limit $T \rightarrow 0$, $\lambda (0)/ \lambda_L(0) = 3.1$. This value can easily be verified by the corresponding entry in table \ref{tab:table1} in reference \cite{Miller1959}. Consequently, $\lambda(0)= 121$~nm. The model curve is in good agreement with the experimental one calculated with $\lambda = 468$~nm (open black circles). We comment on the different values of $\lambda (0)$ in the Discussion section.

The theoretical expression of $\rho_s(T)$ for a conventional type-II superconductor (in the London limit) is given by~\cite{Poole2007,Mitra2017}

\begin{equation}
\rho_s = 1 + \int^{\infty}_{-\infty} \frac{\partial f}{\partial E} d\epsilon,
\end{equation}

with the Fermi function $f = 1/ (1 + \exp(E/k_B T))$ and the Bogoliubov quasiparticle energy $E = (\Delta(T)^2 + \epsilon^2)^{1/2}$. $\Delta(T)$ can be obtained from the gap interpolation formula given by~\cite{Gross1986}

\begin{equation}\label{eq:inter}
\Delta(T) = \Delta(0) \textrm{tanh}\bigg[ \frac{\pi k_B T_c}{\Delta(0)} \sqrt{a\Big(\frac{\Delta C}{\gamma T_c} \Big)\Big(\frac{T_c}{T} -1 \Big)} \bigg],
\end{equation}

where $a = 2/3$. We computed $\rho_s(T)$ for crystal s1 (with steps removed and $H_{ac} \perp c$) with help of equations \ref{eq:nrho} and \ref{eq:inter}, where we inserted $\Delta(0)/k_BT_c = 1.77$ (see table \ref{tab:table2}), and the corresponding $\Delta C / \gamma T_c$-value of 1.45~\cite{Carbotte1990}. The computed theoretical curve is shown as the black solid line in figure \ref{fig:figure5}a. The model curve for type-II superconductivity considerably deviates from the experimental one for crystal s1 (black open circles), which is to be expected if superconductivity in PdTe$_2$ is type-I. On the other hand, a reasonable match between the computed curve and the experimental one can be obtained up to $T/T_c \sim 0.5$ when we reduce $\lambda(0)$ to about 350~nm in equation~5. However, in this case the mismatch at higher temperatures is large. The analysis of the superfluid density of crystals s2 and s3 yields very similar results compared to crystal s1, as demonstrated by the computed $\rho_s$-curves shown in figure \ref{fig:figure5}b,c.

\section{Discussion}

The BCS fits of $\Delta \lambda_a (T)$, presented in figures \ref{fig:figure3} and \ref{fig:figure4}, are consistent with PdTe$_2$ being a conventional $s$-wave superconductor. The ratio $\Delta(0)$/$k_BT_c$ = 1.77 to 1.97 (table \ref{tab:table2} points to weak- or moderately-coupling superconductivity, in agreement with specific heat data~\cite{Amit2018}. The value of the penetration depth $\lambda_a (0)$ extracted from these fits is equal to 468~nm and 377~nm for crystal s1 measured in the perpendicular and parallel configuration, respectively, and largely exceeds the London penetration depth $\lambda_L (0) = 39$~nm. This difference can partly be accounted for by considering non-local electrodynamics, which results in $\lambda(0) = 121$~nm as calculated for a mean free path $\ell= 531$~nm and BCS coherence length $\xi_0=$1.8 $\mu$m. The values $\lambda_L (0) = 39$~nm and $\lambda (0) = 121$~nm fall well within the range of $\lambda$-values of type-I superconductors~\cite{Poole2007}. A small value of $\lambda$ is also in line with a Ginzburg-Landau parameter $\kappa = \lambda / \xi_0 < 1/\surd2$. With the above value $\lambda (0)=468$ nm and $\xi_0$ we calculate $\kappa \approx 0.29$. Type-I superconductivity is furthermore supported by the computed superfluid density in the non-local scenario, which matches the experimental data better than the type-II London model (see figure \ref{fig:figure5}).

At the moment we cannot reconcile the different values of $\lambda (0)$. A possible source of error in the calculation of $\lambda (0)$ via $\ell$ and $\xi_0$ is the assumption of a spherical Fermi surface, which is obviously an oversimplification~\cite{Dunsworth1975}. Another uncertainty is the value of the carrier density, $n$, which was measured on a bulk crystal. Possibly, $n$ varies in different pieces cut from the crystal because of the semimetallic nature of PdTe$_2$ and sample inhomogeneities. A highly relevant experimental technique to determine the absolute value of $\lambda (0)$ in another way is transverse field muon spin rotation ($\mu$SR)~\cite{Blundell1999}. However, this method is suitable for type-II superconductors only, because in a type-I superconductor the magnetic induction is zero (Meissner state) and muon spin precession is absent. While preparing our manuscript other measurements of the magnetic penetration depth of PdTe$_2$ by the TDO technique were reported~\cite{Teknowijoyo2018}. In this work two crystals were measured in the parallel configuration ($H_{ac} \parallel c-$axis). By fitting the BCS expression to $\Delta \lambda$, $\lambda(0)$-values of 220 and 240~nm were derived. These values are about a factor two lower than the values we reported in table \ref{tab:table2}. We remark in reference~\cite{Teknowijoyo2018}, $\Delta \lambda (0)$ was fitted to a power law temperature variation, $T^n$, to compare with predictions for an unconventional superconducting gap function. The authors report good fits with $n=4.2-4.3$, but also remark the data follow the exponential BCS expression (equation \ref{eq:bcsfit}) equally well.

The appearance of small steps in $\Delta \lambda_a$ as shown in figure \ref{fig:figure2} is striking. They are predominantly observed in crystals s1, s2 and s4, but not in s3. Note that crystal s4 was prepared by the Scotch technique by thinning crystal s3. They appear throughout the whole temperature range below $T_c$ and are reproducible. The steps have different sizes up to 3~nm, which is several times the $c$-axis lattice parameter, 0.513~nm~\cite{Lyons1976}. A few larger steps appeared as well. We made a histogram of the step sizes, but did not find a clear periodicity with respect to the thickness of one PdTe$_2$ layer. Possibly microfractures, dislocations and local deformations in the crystal cutting or thinning process lead to a locally lower $T_c$. This in turn might cause  "delayed" superconducting transitions of (parts of) multiple PdTe$_2$ layers. Crystal s1 was measured for $H_{ac} \parallel c$ as well. In this configuration the steps are weaker and occur less frequently. We remark steps are absent in measurements on other materials made with the same experimental set-up and are not an artefact of the measurement method~\cite{Diener2008}. Also no steps were reported in the measurements on PdTe$_2$ reported in reference~\cite{Teknowijoyo2018}.

Besides bulk superconductivity below $T_c = 1.64$~K, the magnetic measurements revealed superconductivity of the surface layer in applied fields below $T_c^S = 1.33$~K~\cite{Leng2017}. We have carefully inspected $\lambda (T)$ in the vicinity of $T_c^S$ but its variation is smooth and no change is detected when cooling the crystals below $T_c^S$. Measurements of $\lambda (T)$ in applied fields may be conducted to shed light on this unusual type of surface superconductivity.

\section{Summary}

We have reported measurements of the relative magnetic penetration depth, $\Delta \lambda (T)$, in the superconducting state ($T_c =1.6$~K) of the Dirac semimetal PdTe$_2$ using the tunnel diode oscillator technique. The experiments have been performed on several crystals and for driving fields $H_{ac} \parallel c$ and $\perp c$. In all cases the temperature variation $\Delta \lambda (T)$ follows an exponential function for $T/T_c < 0.4$, which is consistent with a fully-gapped superconducting state and weak or moderately coupling superconductivity. By fitting the data we extract a magnetic penetration depth $\lambda (0)$ of $\sim 500$~nm for $T \rightarrow 0$. We compare the experimental normalized superfluid density $\rho _s$ to the computed curve for type-I superconductivity with nonlocal electrodynamics and the theoretical expression for type-II superconductivity in the London limit. In the nonlocal electrodynamics model we obtain $\lambda(0) = 121$~nm, with as input parameters for our crystal the London penetration depth $\lambda _L (0) =39$~nm, a mean free path $\ell= 531$~nm and the BCS coherence length $\xi_0=~1.8~\mu$m. We conclude the nonlocal model of type-I superconductivity gives a better match to the experimental superfluid density. At the moment we cannot reconcile the different values of $\lambda (0)$ extracted from the experimental data and the model. Part of the differences might be attributed to simplifications, such as a spherical Fermi surface, in the determination of the relevant parameters, and to variations in the carrier concentration $n$. Small steps are observed in $\Delta \lambda (T)$, which possibly relates to a locally lower $T_c$ due to crystal defects in the single crystalline sample. A next step in the research is to investigate the magnetic penetration depth in small dc-fields, notably in view of the unusual superconductivity of the surface sheath that was detected in applied fields below $T_c = 1.33$~K~\cite{Leng2017}.

\

Acknowledgements - H. Leng acknowledges the Chinese Scholarship Council for grant 201604910855. This work was part of the research program on Topological Insulators funded by FOM (Dutch Foundation for Fundamental Research on Matter).

\section*{References}

\bibliography{References_PdTe2}

\end{document}